\documentclass[12pt]{article}

\usepackage{graphicx}
\usepackage{amssymb}
\usepackage{amsmath}
\usepackage{pstricks}
\usepackage{amsfonts}
\usepackage{pstricks}
\usepackage{color}
\usepackage{verbatim}

\begin{document}

\begin{titlepage}

\begin{flushright}
arXiv:1008.4769
\end{flushright}
\vskip 2.5cm

\begin{center}
{\Large \bf Limits on the Time Variation of the Fermi Constant $G_{F}$ Based on
Type Ia Supernova Observations}
\end{center}

\vspace{1ex}

\begin{center}
{\large Alejandro Ferrero and Brett Altschul }

\vspace{5mm}
{\sl Department of Physics and Astronomy} \\
{\sl University of South Carolina} \\
{\sl Columbia, SC 29208 USA} \\

\end{center}

\vspace{2.5ex}

\medskip

\centerline{\bf Abstract}

\bigskip

The light curve of a type Ia supernova decays
at a rate set by the $\beta$-decay lifetimes of the $^{56}$Ni and
$^{56}$Co  produced in the explosion. This makes such a light curve sensitive
to the value of the Fermi constant $G_{F}$ at the time of the supernova. Using data
from the CfA Supernova Archive, we measure the dependence of the
light curve decay rate on redshift and place a bound on the time variation of
$G_{F}$ of $\left|\frac{\dot{G}_{F}}{G_{F}}\right|<10^{-9}$ yr$^{-1}$.

\bigskip

\end{titlepage}

\newpage

\section{Introduction}

Finding new fundamental physics beyond the standard model and general relativity has
long been a topic of great interest. One way to identify definitive evidence of new
physics is to look for exotic phenomena that
simply cannot occur in the standard physical
theories. One such exotic possibility that has provoked significant interest is
having time-dependent fundamental
constants~\cite{ref-fischer,ref-peik,ref-tzanavaris,ref-gould,ref-cingoz,ref-blatt,ref-murphy,ref-tobar}.

The possibility of time-varying fundamental constants opens up many new lines of
inquiry. The most commonly studied theory with varying constants is one  
with a time-dependent fine structure constant $\alpha$. A quantum
approach to this subject can be found in~\cite{ref-ferrero1}, while other approaches
treat the problem rather phenomenalistically (e.g.~\cite{ref-flambaum1}). 
Another possibility that has been fairly well studied is a change in the quantum chromodynamics scale
$\Lambda_{QCD}$ and therefore the electron-proton mass ratio. 

If some of the fundamental constants in nature are allowed to change over time,
the effects of these changes might be detected. Searches for such temporal variations
entail measurements of the same experimental observable at different points in time,
to see if there is any change. However, since the changes involved
should be tiny, laboratory experiments searching for them demand very high
precision. Less precision is required if large time scales are
involved; a very slow time variation can be observed if the times used
to detect it are long enough. Thus, cosmological observations, which give us a view of
how physics operated millions or billions of years in the past, seem to be good
candidates to test for time variations in fundamental constants.

The purpose of this paper is to study a possible change in the Fermi
constant $G_{F}$. This quantity is defined to be $G_{F}=
\frac{\sqrt{2}}{8}\frac{g^{2}}{M_{W}^{2}}$, with $g$
the standard model's $SU(2)_{L}$ gauge coupling constant and $M_{W}$ the mass of the
$W$ boson. Since the vector boson mass is $M_{W}=\frac{1}{2}gv$, $G_{F}$ is really a measure of
$v$, the Higgs field vacuum expectation value.
The masses of fundamental fermions depend on $v$ and the values
of the fermion-Higgs Yukawa couplings $\lambda_f$, with $m_f=\lambda_f v$. 
So a changing value of $G_F$ would also affect the fermion masses unless the
couplings $\lambda_f$ also changed to compensate. Since there is no evidence of any time variation
in $m_e$ in atomic spectra, we shall assume that such cancellation between the time
variation of $v$ and $\lambda_f$ occurs, even though that is not a particularly natural scenario.

$G_{F}$ also governs the strength of
charged current weak interactions at low energies.
It can be measured with relatively high precision by studying the
lifetimes of muons, and its accepted value is 
$G_{F}=(1.6637\pm0.0001)\times 10^{-5}$
GeV$^{-2}$~\cite{ref-chitwood,ref-barczyk}.
However, laboratory measurements
of decay rates are not precise enough to offer a particularly promising way of
measuring $\dot{G}_{F}$. On the other hand, the light coming from supernovae
explosions can provide a useful way to study weak interactions that occurred at much
earlier times.

In Section~\ref{sec-SN}, we shall discuss the physics of supernovae and how their
light curves are related to the Fermi coupling constant $G_{F}$.
Section~\ref{sec-method} covers the details of the calculational methodology used to
relate the supernova light curve data to $\dot{G}_{F}$. In Section~\ref{sec-data},
we apply this methodology to data from the Harvard-Smithsonian Center
for Astrophysics (CfA) Supernova
Archive~\cite{ref-riess,ref-jha} and state our results. Section~\ref{sec-concl}
presents conclusions and outlook.

\section{Supernova Structure}
\label{sec-SN}

Supernovae are catastrophic variable stars that, having previously produced energy
through steady-state fusion processes, explode and release huge additional amounts
of energy. The amount of energy released is so vast that even cosmologically
distant supernovae can be easily detected in the sky. Normal stars avoid
gravitational collapse because nuclear fusion processes are
constantly generating energy in their cores, where the
high temperature and pressure enable light nuclei to
overcome their electrostatic repulsion and fuse. A large red giant star may burn a
significant portion of its mass into isotopes of iron, beyond which further fusion is
no longer exothermic. However, smaller stars, without thick external layers to
compress the core to the
high temperatures and pressures needed to produce iron, may cease burning when the
core is mostly carbon and oxygen, without having extracted all of this fuel's
available energy. In either case, the core of the star settles down to become
a white dwarf, supported against gravitational collapse by the degeneracy
pressure of its electrons.

If the mass of a white dwarf star is larger than approximately $1.4M_\odot$
(the Chandrasekhar mass $M_{Ch}$), the electron degeneracy pressure can support the star
against gravitational collapse. However, if the mass of the star grows larger than
this limit, the electron pressure can no longer prevent a collapse. Such
a collapse triggers the release of a tremendous amount of energy and the star
becomes a supernova. Iron white dwarfs at the cores of massive stars collapse to
form even more compact objects---neutron stars or black holes. However, of interest
here are the type Ia supernovae produced from carbon-oxygen white
dwarfs. These stars may accrete mass from (or possibly merge with) their partners
in binary systems, until the white dwarf's mass becomes sufficient to trigger a
supernova. When these stars exceed $M_{Ch}$, they begin to collapse.
However, the collapse phase is short-lived;
as the star contracts, there is an increase in temperature and density in
the core of the star, which allows the ignition of carbon fusion. Within a short
period of time, a huge amount of fusion energy is released, blowing the star apart
and increasing its luminosity to enormous values. (The total luminosity may be that
of millions of ordinary suns).

In this paper, we shall be analyzing the light curves
of type Ia supernovae. The near equality of the masses of the white dwarf progenitors
gives them all similar energy outputs and
makes them extremely important standard candles in cosmology.
Type Ia supernovae all seem to follow a characteristic light curve. (Some corrections to
the light curves are required in order for the population of type Ia supernovae to act as
genuinely standard candles, but these are not important for our present purposes.)
The relative uniformity of this curve structure for different stars also
makes them a very useful tool for studying certain physical properties.

The light curve is a plot of the apparent magnitude $m=-2.512\log\frac{f}{f_{0}}$
of the source versus time; $f$ is the radiation flux from the source, and $f_{0}$ is
a reference value of the flux defining 0 magnitude. Using different kinds
of filters, which cover different parts of the spectrum, several different magnitudes
may be measured for a single source. There is a separate light curve for each band,
but the curves for different colors show similar behavior, which
is again consistent across different sources.

At early times, the brightness of a supernova increases rapidly until it reaches a
maximum, which is followed by two regions in which $m$ decreases linearly
(corresponding to exponential decays in the flux).
Since type Ia supernovae are ultimately powered by uncontrolled nuclear fusion,
they produce large quantities of the heavy elements,
especially those elements with binding energies close to the maximum
value. Specifically, each supernova produces a great deal of the
double magic nucleus $^{56}$Ni. This nucleus decays by electron capture
(EC) to $^{56}$Co with a lifetime of 6.077 d, and this decay provides much
of the energy released by the supernova during its earlier stages. The first
linear region along the light curve corresponds to the period during which
$^{56}$Ni decay is the dominant source of heating. The daughter nucleus
$^{56}$Co also decays by EC to $^{56}$Fe, with a lifetime of 77.27 d. At
later times (in the second, more slowly decaying linear region), it is this decay
that primarily fuels the supernova's luminosity. This region, whose extent is
determined by the decay rates of $^{56}$Ni and $^{56}$Co, can be found between
approximately 35 and 100 days after the maximum magnitude is reached.
During this time, some months after a supernova reaches its peak, the light curve
approximately tracks the quantity of $^{56}$Co still present. The falloff in
the luminosity can therefore be used to measure the lifetime of this
nuclide at the time the supernova occurred. This gives direct information
about the value of several standard model parameters at the time of the explosion's
occurrence, including $G_F$. This was first suggested in ~\cite{scher-sper}, when much less data
was available. 

An EC decay rate depends on fundamental constants in the following
fashion. Being a tree-level weak decay, the rate is naturally proportional to
$G_{F}^{2}$.
However, it also depends on the probability that an $s$-state electron will be found
close enough to the nucleus to be taken up by a proton. The nuclear volume
is proportional to $\Lambda_{QCD}^{-3}$.
The square of the electron wave function in the vicinity of the
nucleus is proportional to $a_{0}^{-3}$, where $a_{0}=\frac{1}{m_{e}e^{2}}$ is
the Bohr radius.
The inverse $\Gamma$ of the mean life of a nucleus that is unstable against EC
is therefore proportional to the combination of fundamental quantities
$G_{F}^{2}\alpha^{3}\left(\frac{m_{e}}{\Lambda_{QCD}}\right)^{3}$. (The rate also
depends on the $Q$-value of the decay, which is further dependent on
$\alpha$ and $\frac{m_{e}}{\Lambda_{QCD}}$, in more complicated fashion.)
However, changes in $\alpha$ and $\frac{m_{e}}{\Lambda_{QCD}}
\propto\frac{m_{e}}{m_{p}}$ are much easier to measure than would
be changes in $G_{F}$. $\alpha$ and $\frac{m_{e}}{m_{p}}$ affect atomic
energy levels directly, so their time derivatives can be constrained by comparing
spectra taken at different times. Cosmological and laboratory measurements have
constrained these dimensionless quantities to vary only by parts in $10^{16}$ or
less per year.

A supernova's light curve and total luminosity are also affected by the
value of the gravitational constant $G_{N}$. The amount of $^{56}$Ni
produced in the explosion is proportional to the total mass of the
collapsing star, $M_{Ch}\propto G_{N}^{-3/2}$~\cite{ref-gaztanaga, ref-riazuelo}. 
Observations of supernovae have been used to place bounds on the
variation of $G_{N}$ at the 
$\left|\frac{\dot{G}_{N}}{G_{N}}\right|\lesssim10^{-11}$ yr$^{-1}$ level.
We choose here to look at a model that neglects variations in $G_{N}$, although an
analysis that considered simultaneously varying $G_{N}$ and $G_{F}$ would
also be interesting. We shall also neglect any possible time dependences of
$\alpha$ and $\frac{m_{e}}{\Lambda_{QCD}}$, in light of the strong bounds
on such time dependences. 
This means there is a simple relationship between $\dot{\Gamma}$ and $\dot{G}_{F}$,
\begin{equation}
\frac{\dot{G}_{F}}{G_{F}}=\frac{1}{2}\frac{\dot{\Gamma}}{\Gamma}.
\end{equation}
Bounds on changes in the $^{56}$Co decay rate may thus
be interpreted as bounds on a change in the Fermi constant.

\section{Method}
\label{sec-method}

In the light curve of a type Ia supernova, the slope of the second linear region
is related to the decay rate of $^{56}$Co.
The main goal of this paper is to compare this slope across different
supernovae of different ages. The slope for each one is related to the Fermi
constant at a particular time. We will make this comparison using data from the
CfA Supernova Archive.

Ideally, one could simply measure the slopes of the supernova light curves in their
$^{56}$Co-dominated regions and then determine whether the slopes are correlated with
the ages of the events. However, several corrections must be made before this can
be done.

The light curves do not give a pure measurement of the $^{56}$Co lifetime;
what is observed is the luminosity of the supernova remnant. The
EC decay is the object's main energy source, but the decay energy is
moderated through the complicated interactions of the expanding gas, so
the luminosity does not track perfectly with the $^{56}$Co decay profile.
The details of the remnant dynamics differ a bit from supernova to supernova.
It is not possible to correct for this lack of complete uniformity in the 
structures, and this will necessarily be a source of random error.

Some of the corrections may be handled more simply, however.
This is the case for the correction arising from the well-known Doppler shift
associated with
the expansion of the Universe. Since the inverse of the mean life,
$\Gamma$, measures how fast $^{56}$Co is decaying (and therefore,
the frequency at which EC processes are occurring), the observed
value of $\Gamma$ will be subject to the same cosmological redshift as any other
frequency. In terms of the red shift $z$, the measured value $\Gamma$ on Earth and
the proper decay rate $\Gamma_{0}$ at the exploding star  are related by
\begin{equation}\label{a1}
\Gamma_{0}=(1+z)\Gamma.
\end{equation}
The physical effect of the redshift is to slow down the decay of the light curve.

We shall denote by $\dot{m}$ the slope of the second linear region of 
each particular light curve, as observed on the Earth. This slope and the $^{56}$Co decay rate
should be directly proportional, so we have
\begin{equation}\label{a2}
\Gamma_{0}=(1+z)\Gamma\equiv A(1+z)\dot m= A\dot{m}_{0},
\end{equation}
where $\dot{m}_{0}$ is the proper decay rate of the supernova light
curve, and $A$ is a proportionality constant related to the complex physics of the explosion.
The quantity $\dot{m}_{0}$ is proportional to the square of the Fermi coupling
constant at the time $t_{0}$ that the explosion occurred,
$\dot{m}_{0}\propto\Gamma_{0}\propto G_{F}^{2}(t_0)$. If $G_{F}$ has changed
over time, we would expect to see a change in the decay rates $\dot{m}_{0}$ 
observed for different supernovae. Comparing
the light curves for supernovae of different ages gives a good way to measure  
a change in the Fermi coupling constant.

The light from a particular supernova is emitted and collected at the times
$t_{0}$ and $t$, respectively. Denoting  the scale factor of the Universe by $a(t)$
and the Hubble constant by $H$, we have the standard relations
\begin{eqnarray}\label{a4}
z & = & \frac{a(t)}{a(t_{0})}-1 \\
\label{eq-H}
H & = & \frac{\dot{a}}{a}.
\end{eqnarray}
The factor $H$ converts derivatives with respect to $z$ into ones with respect to
time.
Based solely on the relations (\ref{a4}) and (\ref{eq-H}), one may be tempted to
conclude that 
$\frac{\dot\Gamma_0}{\Gamma_0}=-\frac{H}{\langle\dot m_0\rangle}\frac{d\dot m_0}{dz}$
(where $\langle\dot m_0\rangle$ is the average value of $\dot m_0$ as computed
from the different sources). However,
a new complication arises because of the differences in the behaviors of the
$\dot m_0$ 
observed in different wavelength bands, and this relation for
$\frac{\dot{\Gamma}_{0}}{\Gamma_{0}}$
is subject to further corrections. 
Looking at the light from different color regions, we see
slightly different slopes associated with the different bands. At first glance this
might not seem to be a problem, 
since the value of the proportionality constant $A$ defined in eq. (\ref{a2}) could just be
redefined for each band. The value $\frac{\dot\Gamma_0}{\Gamma_0}$ does not
depend on $A$, so it would not affect the final result.

Nevertheless, the fact that $A$ does not remain constant
across the various bands introduces a new correction. To understand
why a further correction is necessary, let us consider light that is 
detected on Earth using a red filter. This light has been redshifted 
by the expansion of the Universe. If the slopes for the different bands were the
same this would not be a problem, but the red light that is detected
actually had a shorter wavelength when it was emitted during the explosion.
All the red light that is detected on Earth was some amount bluer at the time of
its emission; and we should ideally be comparing light from different sources that
had the same wavelength at the time of emission---not at the time of observation.
If the light curve steepens with increasing wavevector $k$, the redshift would
transform this into a steepening (within a given wavelength band) as a function of
$z$, because the older light originally had a higher frequency. What is required is
similar to a light curve $K$ correction~\cite{ref-hubble}. However, rather than using tabulated
$K$ corrections, we shall, for this low redshift sample of supernovae, calculate
our correction directly from the data set. This is justified because the
technique minimizes our reliance on the results of unrelated fits and because we
shall find that the correction it produces is relatively small.

To account for the redshift's effect, we 
suppose that the dependence of $\dot m_0$ on $k$ is weak, 
and make a linear approximation
\begin{equation}\label{a4a}
(1+z)\dot m_X=(\dot m_X\!)_0+\frac{d\dot m_0}{dk}(k_0-k_X)=(\dot m_X)_0+\frac{d\dot m_0}{dk}(zk_X).
\end{equation}
$\dot m_X$ is the observed slope in the band $X$ with mean wavenumber $k_X$.
Once corrected by $(1+z)$, $\dot{m}_{X}$ gives the proper decay rate $\dot m_0$,
but for a band 
centered at a higher wavenumber $k_0=(1+z)k_X$. The $z$ dependence of $k_0$ generates
a spurious dependence on $z$ in the observational data, which we
must subtract away.

Unlike the simple $(1+z)$ factor relating $\dot m$ to $\dot m_0$ (which is an
exact correction), there is
no particular reason for this effect to be linear in $z$. The wavelength dependence
of the proportionality constant $A$ is a consequence of the complicated dynamics of
the supernova remnants; its detailed form is not known.
If $\frac{d\dot m_0}{dk}$
is itself a function of $z$, there will be $\mathcal O(z^2)$ corrections to (\ref{a4a}).
At the same order, it would also be necessary to correct the average $\langle(\dot m_X)_0\rangle$
to account for the fact that we are actually averaging decay rates in different color regions.
We are neglecting these possibilities (which would correspond to corrections that are
higher order in $z$ in our final formula for $\frac{\dot{\Gamma}_{0}}{\Gamma_{0}}$),
because the redshifts in the sample we shall study
range only from $3.9\times 10^{-3}$ to $2.4\times 10^{-2}$. However, these corrections can be 
computed, and working with sources at higher $z$ would require us to have a better
knowledge of how $\dot m_0$
depends on $k$. Yet there is a limit to the precision that can be derived from more detailed
models, since there are data available for only a handful of wavelength
bands (five in this case), and data points at multiple frequencies would be required
to map out the detailed dependence of $A$ on $k$.

In principle, we could try to correct the values of $\dot m_0$ for each supernova
individually, by studying the dependence of its light curves on $k$.
We could then use the corrected values obtained separated for each source
to compute $\frac{1}{\langle\dot m_0\rangle}\frac{d\dot m_0}{dz}$. 
However, this would require additional fitting for each supernova, 
which may introduce additional errors and which cannot be practically implemented for many of
the sources because of the limited data available. We
shall therefore make only a single calculation of
$\frac{d\langle\dot m_0\rangle}{dk}$, using the combined data from all the sources.

Using eq. (\ref{a4a}) we find our final formula for $\frac{\dot\Gamma_0}{\Gamma_0}$,
\begin{equation}\label{a4c}
\frac{\dot\Gamma_0}{\Gamma_0}=-\frac{H}{\langle\dot m_0\rangle}\left[\frac{d\dot m_0}{dz}
-\frac{2\pi}{\lambda}\left(\frac{d\langle\dot m_0\rangle}{dk}\right)\right],
\end{equation} 
including the linear correction to eliminate the spurious $z$ dependence caused by a
nonzero $\frac{d\dot m_0}{dk}$. Evaluating this requires multiple fits. The order of the
fitting procedures
is as follows. First, $\dot m_0$ is determined for each supernova in each band separately. 
These are then fitted as a linear function of the redshift $z$,
giving both an average $\langle\dot m_0\rangle$ and a slope $\frac{d\dot m_0}{dz}$ for each band. 
From the fit of the five average values $\langle\dot m_0\rangle$
versus $k$, the parameter $\frac{d\langle\dot m_0\rangle}{dk}$ can be extracted.
Then $\frac{\dot{\Gamma}_{0}}{\Gamma_{0}}$ can be evaluated according to
eq.~(\ref{a4a}); this evaluation can be done separately in each of the five bands,
although the single fitted value for the quantity
$\frac{d\langle\dot{m}_{0}\rangle}{dk}$ is common to all of them.

The overall negative sign in eq.~(\ref{a4c}) merits some explanation. Because the light curves are
decaying in the region of interest, $\dot{m}_{0}$ is always negative. If $\dot{m}_{0}$ is
more negative at greater values of $z$ ($\frac{d\dot{m}_{0}}{dz}<0$ and
$\frac{1}{\langle \dot{m}_{0}\rangle}\frac{d\dot{m}_{0}}{dz}>0)$, the decay rate was
faster at earlier times. This means $\Gamma_{0}$ and thus $G_{F}$ are decreasing as
time passes, leading to the negative sign in eq.~(\ref{a4c}).

\section{Data Analysis}
\label{sec-data}


\begin{figure}
\begin{center}
\includegraphics[scale=1.2]{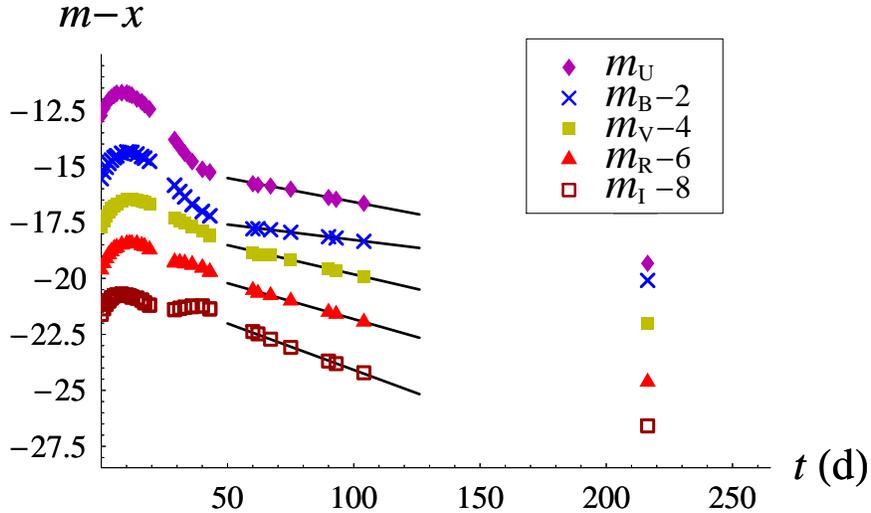}
\caption{Light curves for SN 1998aq. The magnitudes for the different
bands are $m_U$, $m_B-2$, $m_V-4$, $m_R-6$,
and $m_I-8$. The black lines represent linear fits between
50 and 130 days after the peak.}
\label{f1}
\end{center}
\end{figure}



Using the method we have discussed, we can now place a bound on the time
variation rate of $G_{F}$ using data from the CfA Supernova Archive.
For purposes of illustration, we shall initially focus on one particular
supernova, SN 1998aq~\cite{ref-riess2}, for which $z=3.95\times10^{-3}$.
Its light curves are shown in fig.~\ref{f1}, plotting the supernova's magnitude as
a function of time $t$ in days.
The linear behavior arising from the $^{56}$Co decay
can be seen in a region covering approximately 50--130 days after the maximum magnitude
was recorded.
Although the data has small errors related to the measurements, these
have been omitted for clarity of presentation. To find the slopes in the different bands, we used a weighted linear fit in the region of interest, minimizing
the $\chi^2$ per degree of freedom. The results of these fits are also shown in
fig.~\ref{f1} and the fitted slopes are listed in table~\ref{t1}. In order to
transform these results into the reference frame of supernova, we just use the
relation $\dot{m}_{0}=(1+z)\dot{m}$.

\begin{table}
\begin{center}
\begin{tabular}{|c|c|}
\hline
\rule{0pt}{2.4ex}\!\!
Band &  $\dot{m}$ (10$^{-4}$\,d$^{-1}$)\\
\hline
\rule{0pt}{2.5ex}\!\!
  U  &  $-214\pm7$\\
  B  &  $-138\pm2$ \\
  V  &  $-261\pm4$\\
  R  &  $-321\pm4$\\
  I  &  $-415\pm6$\\
\hline
\end{tabular}
\end{center}
\caption{Fitted values of the slope $\dot{m}$ for SN1998aq, in units of $10^{-4}$
d$^{-1}$.}
\label{t1}
\end{table}

We have applied the same procedure to multiple sources. Not all sources in the
CfA Supernova Archive had sufficient data to be fitted reliably, and not all bands
were represented for all sources.
In table \ref{t2}, we summarize the results found for the different 
supernovae we analyzed. 
The slopes $\dot{m}$ are in the observer reference frame on Earth (no $1+z$
factor included). There was one further complication associated with fitting some
of the light curves. For some sources, there is limited data at early times, and
it was not possible to resolve the date at which the light curve achieved its
absolute maximum. In many cases, we could identify the proper linear region by
inspection, but if there was any doubt whether a given data point should be 
included, the fitted region was chosen so as to produce the best linear fit.
\begin{table}[h]
\begin{center}
\begin{scriptsize}
\begin{tabular}{|l|r|l|l|l|l|l|}
\hline
\rule{0pt}{2.3ex}
Source   &        $ 10^5\,z $         & $\dot m_U$ ($10^{-4}\,$d$^{-1}$) & $\dot m_B$ ($10^{-4}\,$d$^{-1}$)
& $\dot{m}_V$ ($10^{-4}\,$d$^{-1}$) & $\dot m_R$ ($10^{-4}\,$d$^{-1}$) & $\dot m_I$ ($10^{-4}\,$d$^{-1}$) \\
\hline
\rule{0pt}{2.3ex}\!\!
1995bd& $160 $ & \hspace{0.95cm}--         & \hspace{0.3cm}$-171\pm7 $ & \hspace{0.3cm}$-271\pm3 $ & \hspace{0.95cm}--         & \hspace{0.95cm}--          \\
1998aq& $395 $ & \hspace{0.3cm}$-214\pm8 $ & \hspace{0.3cm}$-138\pm2 $ & \hspace{0.3cm}$-261\pm4 $ & \hspace{0.3cm}$-321\pm4 $ & \hspace{0.3cm}$-415\pm6  $ \\
1994ae& $427 $ & \hspace{0.95cm}--         & \hspace{0.3cm}$-145\pm2 $ & \hspace{0.3cm}$-224\pm4 $ & \hspace{0.3cm}$-286\pm7 $ & \hspace{0.3cm}$-359\pm8  $ \\
1995al& $514 $ & \hspace{0.95cm}--         & \hspace{0.3cm}$-152\pm8 $ & \hspace{0.3cm}$-246\pm4 $ & \hspace{0.3cm}$-311\pm3 $ & \hspace{0.3cm}$-366\pm1  $ \\ 
1995D & $654 $ & \hspace{0.95cm}--         & \hspace{0.3cm}$-131\pm5 $ & \hspace{0.3cm}$-248\pm6 $ & \hspace{0.3cm}$-333\pm35$ & \hspace{0.3cm}$-384\pm24 $ \\
1999gh& $767 $ & \hspace{0.3cm}$-182\pm20$ & \hspace{0.3cm}$-175\pm5 $ & \hspace{0.3cm}$-251\pm5 $ & \hspace{0.3cm}$-322\pm12$ & \hspace{0.3cm}$-299\pm33 $ \\
2000cx& $793 $ & \hspace{0.3cm}$-226\pm4 $ & \hspace{0.3cm}$-187\pm2 $ & \hspace{0.3cm}$-257\pm8 $ & \hspace{0.3cm}$-312\pm8 $ & \hspace{0.3cm}$-347\pm19 $ \\
1999ac& $949 $ & \hspace{0.3cm}$-240\pm6 $ & \hspace{0.3cm}$-175\pm14$ & \hspace{0.3cm}$-231\pm11$ & \hspace{0.3cm}$-313\pm13$ & \hspace{0.3cm}$-317\pm40 $ \\
1998bp& $1042$ & \hspace{0.95cm}--         & \hspace{0.3cm}$-187\pm13$ & \hspace{0.95cm}--         & \hspace{0.95cm}--         & \hspace{0.3cm}$-289\pm42 $ \\
1998es& $1056$ & \hspace{0.95cm}--         & \hspace{0.3cm}$-187\pm16$ & \hspace{0.3cm}$-314\pm15$ & \hspace{0.3cm}$-376\pm7 $ & \hspace{0.95cm}--          \\
1995E & $1156$ & \hspace{0.95cm}--         & \hspace{0.95cm}--         & \hspace{0.3cm}$-253\pm20$ & \hspace{0.3cm}$-321\pm17$ & \hspace{0.3cm}$-441\pm23 $ \\
1999dq& $1432$ & \hspace{0.3cm}$-254\pm1 $ & \hspace{0.3cm}$-163\pm15$ & \hspace{0.3cm}$-250\pm17$ & \hspace{0.3cm}$-332\pm18$ & \hspace{0.3cm}$-387\pm33 $ \\  
1999aa& $1443$ & \hspace{0.3cm}$-210\pm2 $ & \hspace{0.3cm}$-130\pm8 $ & \hspace{0.3cm}$-248\pm7 $ & \hspace{0.3cm}$-316\pm6 $ & \hspace{0.3cm}$-414\pm7  $ \\
1993ae& $1904$ & \hspace{0.95cm}--         & \hspace{0.95cm}--         & \hspace{0.3cm}$-231\pm41$ & \hspace{0.3cm}$-316\pm24$ & \hspace{0.3cm}$-428\pm123$ \\
1994M & $2296$ & \hspace{0.95cm}--         & \hspace{0.3cm}$-149\pm7 $ & \hspace{0.3cm}$-294\pm18$ & \hspace{0.3cm}$-399\pm2 $ & \hspace{0.3cm}$-454\pm1  $ \\
1995ak& $2300$ & \hspace{0.95cm}--         & \hspace{0.3cm}$-243\pm20$ & \hspace{0.3cm}$-310\pm12$ & \hspace{0.3cm}$-418\pm25$ & \hspace{0.3cm}$-489\pm57 $ \\
\hline
\end{tabular}
\end{scriptsize}
\end{center}
\caption{Fitted values of $\dot{m}$, in units of $10^{-4}$ d$^{-1}$. Sources and
bands for which the data were insufficient are marked ``--''.}
\label{t2}
\end{table}

The data available for some supernovae and some wavelength bands were not of
sufficient quality to be useful. We omitted those sources for which the standard
error in the fit of $\dot{m}$ was larger than 0.015 d$^{-1}$ and those with fewer
than three data points in the region of interest. The former was indicative of
substantial deviations from linear behavior (that is, nonexponential decay in
the luminosity), and the later simply represented too few points to be useful.

\begin{figure}
\begin{minipage}[b]{0.5\linewidth}
\centering
\includegraphics[scale=0.6]{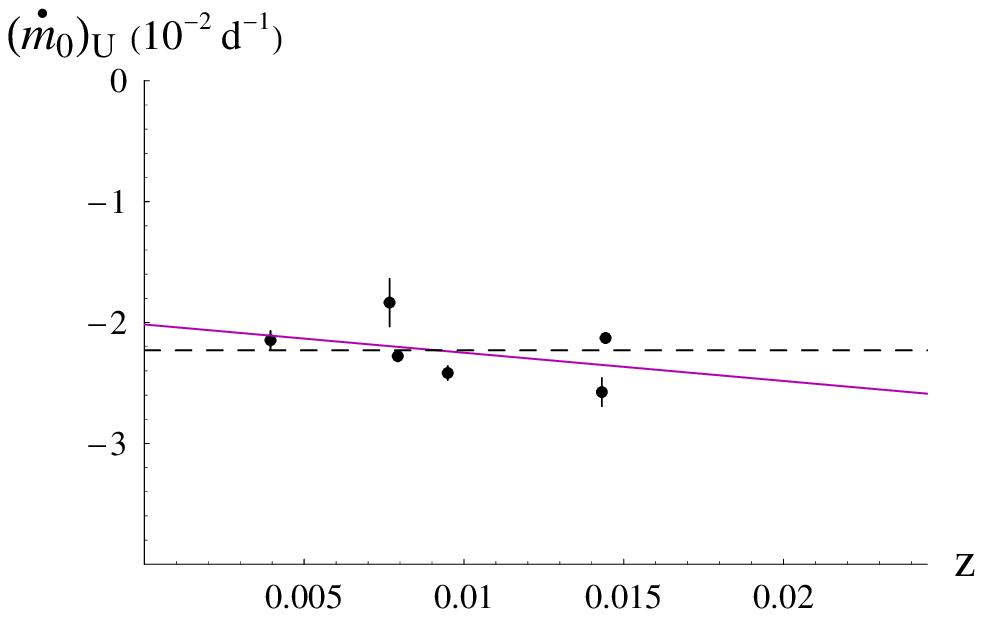}
\end{minipage}
\hspace{0.5cm}
\begin{minipage}[b]{0.5\linewidth}
\centering
\includegraphics[scale=0.6]{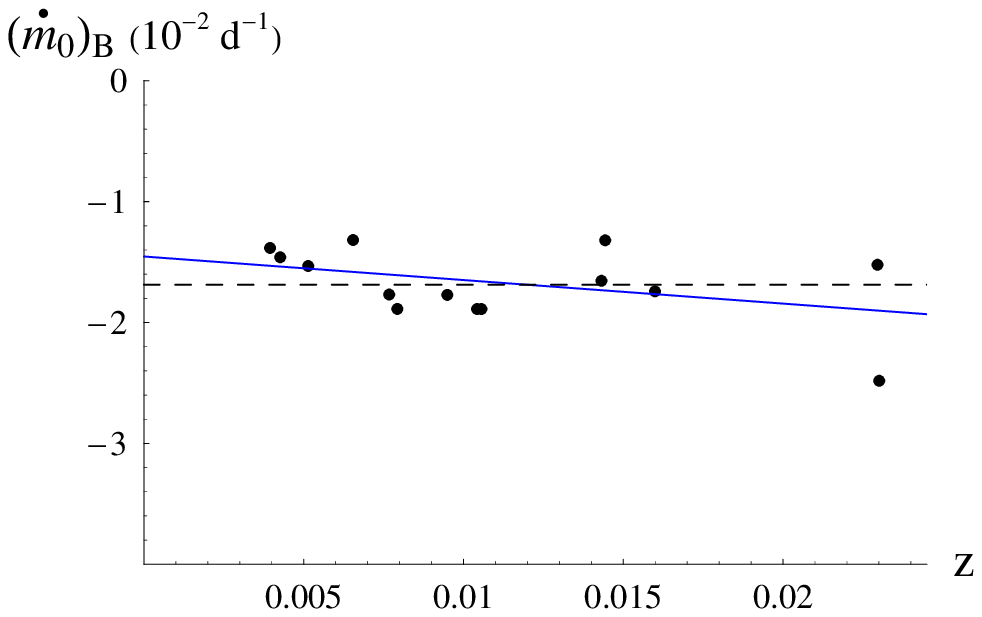}
\end{minipage}
\begin{minipage}[b]{0.5\linewidth}
\centering
\includegraphics[scale=0.6]{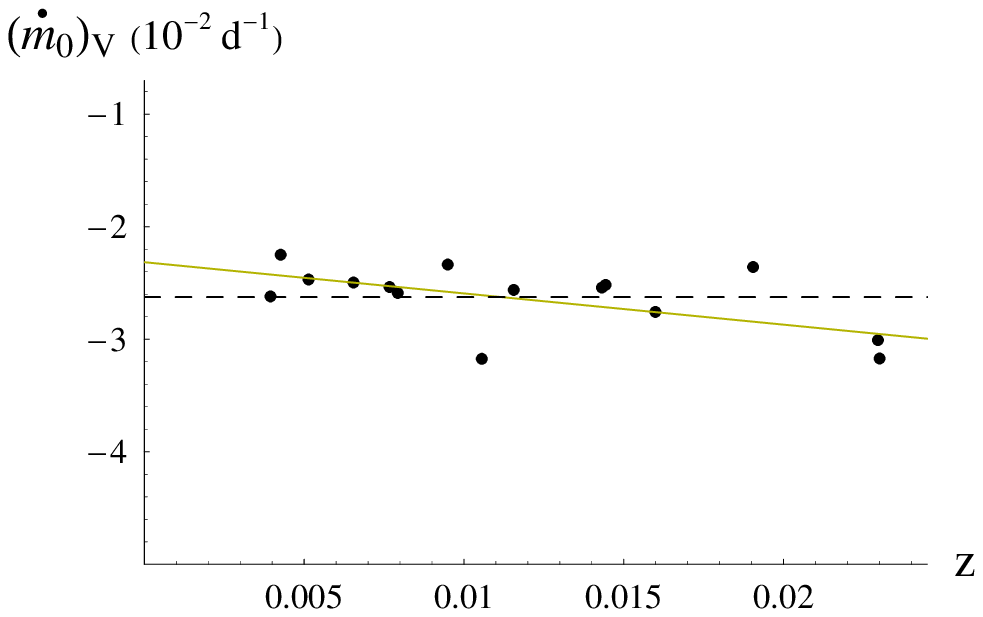}
\end{minipage}
\hspace{0.5cm}
\begin{minipage}[b]{0.5\linewidth}
\centering
\includegraphics[scale=0.6]{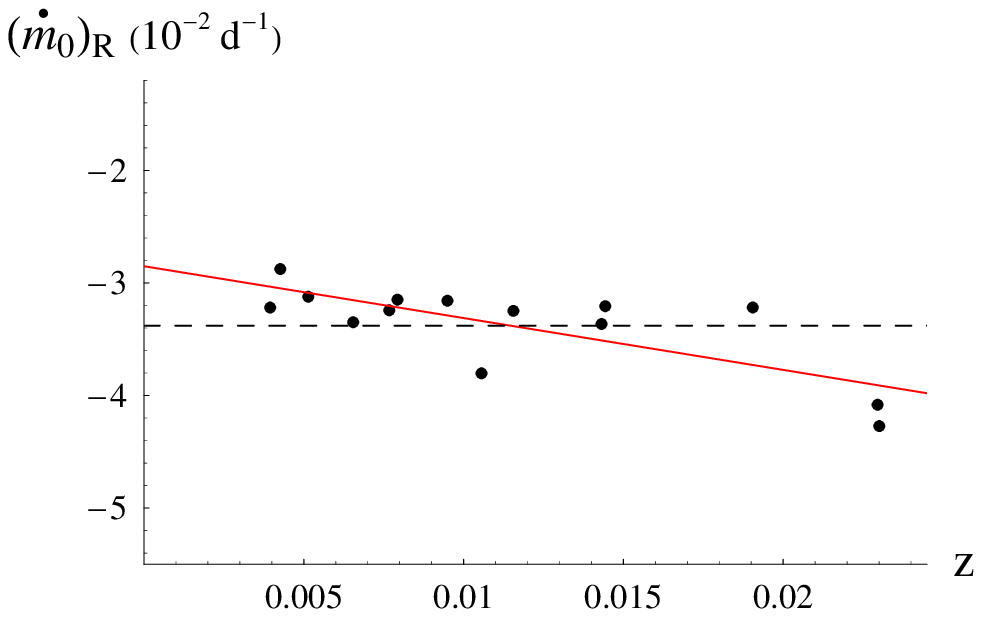}
\end{minipage}
\begin{minipage}[b]{0.5\linewidth}
\centering
\includegraphics[scale=0.6]{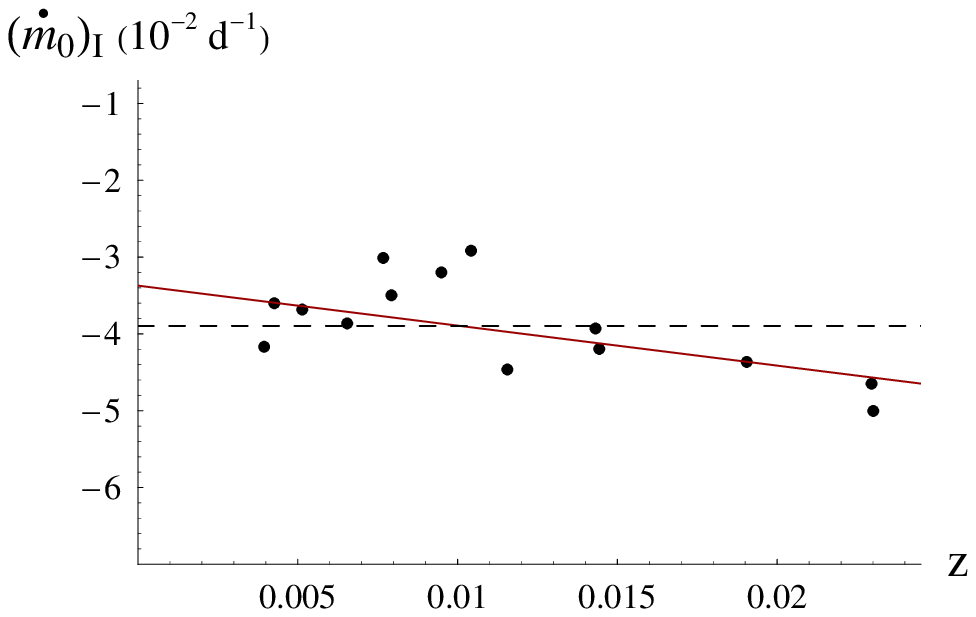}
\end{minipage}
\caption{Plots of $\dot{m}_{0}$ versus $z$ for the U, B, V, R, and I bands
(denoted by colors as in fig.~\ref{f1}). In each plot, the dashed horizontal
line marks the mean $\langle\dot{m}_{0}\rangle$; the substantial difference in
this mean for different wavelength regions indicates a nonzero
$\frac{d\langle\dot{m}_{0}\rangle}{dk}$.}
\label{f2}
\end{figure}

The $z$ dependence of $\dot{m}_{0}$ for each wavelength band is shown in
fig.~\ref{f2}. It is evident that $\frac{d\dot{m}_{0}}{dz}$ is negative in each of
the five
distinct wavelength regimes. The values of the slopes, as well as some other
relevant quantities, are collected in table~\ref{t5}.


\begin{table}[h]
\begin{center}
\begin{tabular}{|c|c|c|c|c|}
\hline
\rule{0pt}{2.8ex}
Band & $\lambda$ (nm)& $\frac{d\dot m_0}{dz}$ ($10^{-2}\,$d$^{-1}$) & $\langle\dot m_0\rangle$ ($10^{-3}\,$d$^{-1}$)\\
\hline
\rule{0pt}{2.7ex}\!\!
 U   & $ 350.0 $ & $ -23\pm27$  & $ -22\pm3 $ \\
 B   & $ 438.0 $ & $ -19\pm12$  & $ -17\pm3 $ \\
 V   & $ 546.5 $ & $ -28\pm10$  & $ -26\pm3 $ \\
 R   & $ 647.0 $ & $ -46\pm11$  & $ -34\pm4 $ \\
 I   & $ 786.5 $ & $ -52\pm19$  & $ -39\pm7 $ \\
\hline 
\end{tabular}
\end{center}
\caption{Fitted means and $z$-dependences of $\dot{m}_{0}$ for the five
wavelength bands. $\lambda$ is the central wavelength for each band.}
\label{t5}
\end{table}

The last fit required was for the wavelength dependence of $\langle\dot m_0\rangle$.
The result was $2\pi\frac{d\langle\dot m_0\rangle}{dk}=(9.3\pm4.8)$ nm/d.
The correction associated with this factor will make only a small contribution to the
final value of $\frac{\dot\Gamma_0}{\Gamma_0}$. This correction, along with the other
factors necessary to produce the final computed value of
$\frac{\dot{\Gamma}_{0}}{\Gamma_{0}}$, are given in table~\ref{t1a}. The fourth
column of this table collects all the corrections for each band, giving the full
$z$-dependence of the $^{56}$Co decay rate $\Gamma_{0}$; only a factor
of the Hubble constant $H$ is needed to convert this into a time dependence.

\begin{figure}
\centering
\includegraphics[scale=0.65]{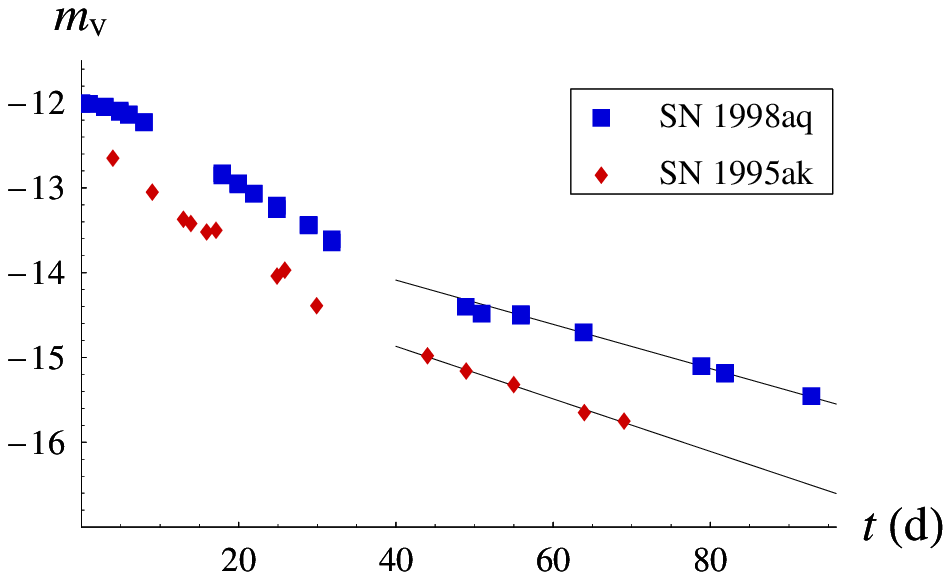}
\caption{V-band light curves of SN 1995ak and SN 1998aq, showing
linear fits in the $^{56}$Co decay region. 
The maximum intensity occurs at $t=0$, where $m_\textrm V$ is normalized to 
$-12.5$ and $-12$, respectively.}
\label{f3}
\end{figure}

\begin{table}[h]
\begin{center}
\begin{tabular}{|c|c|c|c|}
\hline
\rule{0pt}{3ex}
Band & $-\frac{1}{\langle\dot m_0\rangle}\frac{d\dot m_0}{dz}$ & 
$\frac{k}{\langle\dot m_0\rangle}\frac{d\langle\dot m_0\rangle}{dk}$ 
& $\frac{1}{H}\frac{\dot\Gamma_0}{\Gamma_0}$ \\   
\hline
\rule{0pt}{2.5ex}\!\!
 U   & $ -10.5\pm12.1 $ & $ -1.2\pm0.6 $ & $ -11.7\pm12.1 $ \\
 B   & $ -11.5\pm7.7  $ & $ -1.3\pm0.7 $ & $ -12.8\pm7.8 $ \\
 V   & $ -10.6\pm4.0  $ & $ -0.7\pm0.3 $ & $ -11.2\pm4.0 $ \\
 R   & $ -13.6\pm3.8  $ & $ -0.4\pm0.2 $ & $ -14.1\pm3.8 $ \\
 I   & $ -13.4\pm5.5  $ & $ -0.3\pm0.2 $ & $ -13.7\pm5.5 $  \\
 \hline
\end{tabular}
\end{center}
\caption{Contributions to $\frac{\dot\Gamma_0}{\Gamma_0}$ using supernovae in the domain
$3.9\times 10^{-3}<z<2.4\times 10^{-2}$.}
\label{t1a}
\end{table}

Using the values from table~\ref{t1a}, we may compute the weighted average of
$\frac{\dot{\Gamma}_{0}}{\Gamma_{0}}$ for all the frequency bands.
The result is $\frac{1}{H}\frac{\dot\Gamma_0}{\Gamma_0}=-12.7\pm7.4$, so
\begin{equation}\label{a11}
\frac{\dot{G}_{F}}{G_{F}}=(-4.8\pm2.8)\times 10^{-10}\,\textrm{yr}^{-1}.
\end{equation}
where we have used $H=(74.2\pm3.6)$ km/s/Mpc~\cite{ref-riess3}, the uncertainty in
which
introduces a small additional error.

\section{Conclusions}
\label{sec-concl}

Our main result, eq.~(\ref{a11}), shows that the Fermi constant $G_{F}$ may be
decreasing with time; however, the evidence for a change
is at less than the $2\sigma$ level. The error is too large to conclude that 
there is any systematic change in $\Gamma_{0}$ or $G_{F}$. Stating our results
as a $2\sigma$ bound, we have
\begin{equation}
\left|\frac{\dot{G}_{F}}{G_{F}}\right|<10^{-9}\,\textrm{yr}^{-1}.
\end{equation}

The main source of
error was the fitting errors associated with the
determination of $\frac{d\dot m_0}{dz}$. 
Nevertheless, the results for the variation of the decay rate of the $^{56}$Co
isotope seem to be roughly consistent across the five different wavelength bands.

Most of the statistical pull toward a
negative value of $\frac{\dot G_F}{G_F}$ comes from two sources, SN1995ak and SN1994M,
at comparatively high redshifts. The V-band light curves for SN 1995ak and the more typical
SN 1998aq are shown in figure \ref{f3}. The two sources at the highest $z$ values show markedly more
rapid light curve decays than do the lower-$z$ supernovae. When sources such as these are used for cosmic distance scale 
measurements, the more rapid decay is removed by applying a stretch factor $s$ 
to the time axis~\cite{ref-stretch}. However, the use of such an empirically determined factor appears to be
inappropriate here; it could rescale away any actual signal from variations in $G_F$ that was
seen from the data. If $G_F$ is not varying, some other systematic difference between these 
sources and the others is probably responsible, but the nature of this difference is unknown.

The second term in eq.~(\ref{a4c}), which comes from the $k$ dependence of $\dot m_0$,
is considerably smaller than the first term, which accounts for its direct dependence on
$z$. Thus, the poorly understood physics that causes the light curve to depend on
wavelength contributes only a small adjustment to the inferred value of
$\frac{\dot{G}_{F}}{G_{F}}$. This was found to be true in all the bands.
Since the leading order correction due to the $k$-dependence is small, as is $z$ for the
sources we have considered, we do not expect there would be any important changes to 
$\frac{\dot\Gamma_0}{\Gamma_0}$ if higher order in $z$ corrections were taken into account.
However, if sources with greater redshifts were to be analyzed, the higher order
corrections would need to be considered and their contributions could be important.

More sources should be analyzed for this method to produce its best possible result
for the time variation of the Fermi constant.
Although the CfA Supernova Archive contained plenty of information for small redshift
supernovae ($z<2.4\times 10^{-2}$), more large redshift
supernovae should also be analyzed in order to obtain better results. At present,
the $10^{-9}$ yr$^{-1}$ bound on the fractional time variation can be seen as
ruling out roughly a 10\% change in $G_{F}$ over about 100 million years of time.
Although the late
time behavior of light curves is more difficult to observe for more distant
supernovae, using older sources could significantly improve these results.
It would also be desirable to have more data with well-defined locations for their
maxima; light curves with more
data points in the linear region of interest would also be particularly valuable.

We found no significant evidence for a change in $G_{F}$ over the cosmological
interval we studied, $3.9\times 10^{-3}< z< 2.4\times10^{-2}$. However, if we
did see a systematic shift in the decay rate of the supernova light
curves, it could probably be more reasonably explained by something other
than a change in the strength of the weak interaction. We have not
measured $G_{F}$ (or the half-life of $^{56}$Co) directly. Changes in supernova
composition over the lifetime of the sample could potentially have
affected the decay rate of the light curve, changing the way the $^{56}$Co
decay products heat the expanding remnant. Statistically significant evidence for
systematic changes in the supernova light curves over the lifetime of the universe would
be suggestive, but it would by no means be compelling evidence for changing fundamental
constants.

This analysis has determined the rate of change of the dimensional quantity
$G_{F}$. In general, only the time dependence of dimensionless numbers can
be defined unambiguously. A quantity with dimensions must be compared with
some similarly dimensioned reference unit, which may itself depend on time
in a correlated fashion; what reference is chosen will affect how the
measured quantity appears to depend on time. However, there is
essentially no problem in this instance. We can calculate how the
dimensional parameter $\Gamma$ (or $G_{F}$) depends on time with no
meaningful ambiguity---just as one can observe the growth in a child's weight over
time without confronting any ambiguity---since any reference
with which the dimensional observable is implicitly compared is
independent of $G_{F}$. All possibly relevant time standards with which
$\Gamma_{0}$ could be compared have the same dependence on $G_{F}$---none.

In order to extend these bounds to data sets with older, more distant supernovae, we
would need to account for several additional complications. For very distant sources
(and especially for late periods along the light curve), the quality of the magnitude
data diminishes. In addition to limiting the amount of usable data, this could
introduce unexpected biases into the data. (For example, light curves with more rapid
$^{56}$Co decay rates might be underrepresented in the set of usable data, because their
luminosity falls below the threshold for detection more quickly, providing too few
data points to map out the relevant region of their light curves accurately.)
A more systematic $K$ correction would probably also be required at high redshifts,
and it would be desirable to make use of more detailed models of type Ia supernovae
structure, so as to get at the $^{56}$Co lifetime more directly.

These are not the first published bounds on $\dot{G}_{F}$. Although the Fermi coupling
constant cannot straightforwardly be observed through its effects on atomic spectra, it
does effect nuclear energy levels in a small way. Aside from astrophysical observations
of ancient events and precision laboratory experiments, the most successful method
for placing bounds on the time variations of fundamental constants has been to look at
the isotopes produced by the natural reactors that operated near Oklo, Gabon about
$2\times 10^{9}$ years ago. A bound on the fraction change in the dimensionless quantity
$\beta=G_F m_p^2$ at the $10^{-11}$ yr$^{-1}$ level was based on
the production of samarium isotopes at Oklo~\cite{ref-damour}. However, bounds based on
the Oklo data require a very detailed understanding of the operation of the reactor, as
well as complicated nuclear structure calculations, whose reliability as regards the
effects of $G_{F}$ is somewhat questionable. Therefore, our technique, which can be
extended to older supernovae, may offer the best possibility for placing robust bounds
on the time variation of the Fermi constant.

\section*{Acknowledgments}
The authors wish to thank Santiago Gonzalez for his helpful comments.
This research has made use of the CfA Supernova Archive, which is funded in part by the
National Science Foundation through Grant No. AST 0606772.

\end{document}